\documentclass[universe,article,accept,moreauthors,pdftex]{Definitions/mdpi} 
\usepackage{upgreek}

\firstpage{1} 
\pubvolume{1}
\issuenum{1}
\articlenumber{5}
\pubyear{2020}
\copyrightyear{2020}
\history{Received: date; Accepted: date; Published: date}

\def\prd{Phys. Rev. D}
\def\prl{Phys. Rev. Lett.}

\Title{Neutron Stars and Dark Matter}

\Author{Antonino Del Popolo $^{1,2,3}$, Morgan Le~Delliou $^{4,5,}$* and Maksym Deliyergiyev $^{6}$}

\AuthorNames{Antonino Del Popolo, Morgan Le~Delliou and Maksym Deliyergiyev}

\address{%
$^{1}$ \quad Dipartimento di Fisica e Astronomia, University of Catania, Viale Andrea Doria 6, 95125 Catania, Italy; adelpopolo@oact.inaf.it\\
$^{2}$ \quad Institute of Astronomy, Russian Academy of Sciences, 119017 Pyatnitskaya str., 48, Moscow, Russia\\
$^{3}$ \quad INFN sezione di Catania, Via S. Sofia 64, I-95123 Catania, Italy\\
$^{4}$ \quad Institute of Theoretical Physics, School of Physical Science and Technology, Lanzhou University, No. 222, South Tianshui Road, Lanzhou, Gansu 730000, China\\
$^{5}$ \quad Instituto de Astrof\'isica e Ci\^encias do Espa\c co, Universidade de Lisboa, Faculdade de Ci\^encias, Ed. C8, Campo Grande, 1769-016 Lisboa, Portugal\\
$^{6}$ \quad Departement de Physique Nucleaire et Corpusculaire, University of Geneva, CH-1211 Geneve 4, Switzerland; maksym.deliyergiyev@unige.ch}

\corres{Correspondence: delliou@lzu.edu.cn or Morgan.LeDelliou.ift@gmail.com}

\abstract{Neutron stars change their structure with accumulation of dark matter. We study how their mass is influenced from the environment. Close to the sun, the dark matter accretion from the neutron star does not have any effect on it. Moving towards the galactic center, the density increase in dark matter results in increased accretion. At distances of some fraction of a parsec, the neutron star acquire enough dark matter to have its structure changed. We show that the neutron star mass decreases going towards the galactic centre, and that dark matter accumulation beyond a critical value collapses the neutron star into a black hole. Calculations cover cases varying the dark matter particle mass, self-interaction strength, and ratio between the pressure of dark matter and ordinary matter. This allow us to constrain the interaction cross section, $\sigma_{\rm dm}$, between nucleons and dark matter particles, as well as the dark matter self-interaction cross section.}

\begin{document}

\section{Introduction}
Explaining structure formation without the need to modify gravity in current cosmological models entails the introduction of dark matter (DM). Such an introduction induces well documented gravitational effects on structures~\citep{Betoule:2014frx,Ade:2013zuv}, however this dominant part of matter continues to elude detection of its constituting particles, whether by direct detection, through accelerators or nuclear recoil experiments~\citep{Chatrchyan:2012me,ATLAS:2012ky,Agnese:2014aze,Angloher:2011uu,Felizardo:2011uw,Klasen:2015uma,Akerib:2013tjd,Ahmed:2010wy,Bernabei:2010mq,Aalseth:2010vx,Aprile:2012nq}, 
or by indirect searches, through scrutinizing WIMP 
annihilation detection~\citep{Conrad:2014tla}, effects on DM stars~\citep{Dai:2009ik,Kouvaris:2015rea} or through other indirect effects such as proposed in~\citep{Bertolami:2007zm,LeDelliou:2007am,Bertolami:2008rz,Bertolami:2007tq,Bertolami:2012yp,Abdalla:2007rd,Abdalla:2009mt,Delliou:2014awa}.

This evasion welcomes proposals for alternate testing strategies of DM effects. Neutron stars (NSs) offer laboratories that can accrete DM in extreme densities environments. The~presence of DM
thus strains the NSs saturated neutron Fermi gas. The~Tolman--Oppenheimer--Volkoff (TOV) equation~\citep{Tolman:1939jz,Oppenheimer:1939ne} governs the amount of
DM that can be acquired by a NS,
as in e.g.,~\citep{Tolos2015}.

In particular, the~heat produced by the annihilation of WIMPs in the DM core should modify the temperature and luminosity of ancient stars~\citep{Kouvaris2008,Bertone2008,Kouvaris2010,deLavallaz2010}. 
Stars older than 10 million years should have a temperature of $\simeq$10$^4$ K~\citep{Kouvaris2008}, while the coldest observed NS have larger temperatures, \mbox{$10^5\text{--}10^6$ K}, and~WIMPs annihilation has fundamentally no practical importance~\citep{Kouvaris2008}. 
The closer NSs are to the galactic center (GC), the~larger their temperature, and~they can reach temperature of $10^6$ K, and~luminosities of $10^{-2}$ L$_{\odot}$~\citep{deLavallaz2010}. 
Observation of such large values are difficult~\citep{Sandin2009} since the typical temperatures produced by the influence of annihilation effects on the NS cooling will give rise to temperatures in the range 3000--10,000 K (depending on DM density). The~black body emission at these temperatures peak in the UV and obscurations due to dust at the galactic centre~\citep{deLavallaz2010} makes it very difficult to have precise measurements of the NS surface temperature. At~the same time, even the detection of a NS with higher temperature than average is 
not a conclusive proof for WIMPs existence since NS could have a larger temperature in its young phase and cools via the Urca process through neutrino emission, or~because of mass accretion from a binary companion.
One alternate DM model from the framework of WIMPs can be proposed, under~the name of asymmetric dark matter (ADM) model. 
That model's present DM abundance proceeds from a similar origin to that of visible matter~\citep{Petraki2013}\footnote{Mirror matter, with~this definition, falls into a peculiar case of thus defined ADM.}. 
As ADM particles do not annihilate, they can thermalize and concentrate in a small radius, allowing for the formation of extremely compact objects, and~modifying the $M-R$ relation. 
\mbox{The properties} of DM and of the equation of state (EoS) of the NS can be constrained by comparing the $M-R$ relation of usual NSs with NSs containing DM~\citep{Ciarcelluti2010}. 
Moreover, for~a larger accumulated DM amount than a critical value~\citep{Kouvaris2013}, the~resulting NS could collapse into a mini black-hole. This allows us to constrain further the cross section and mass of the DM particles~\citep{Bertone2008}.

Accretion of non-annihilating DM on NSs adds to other known properties, connected to non-annihilation~\citep{Li:2012ii,Sandin2009,Leung:2011zz,Xiang:2013xwa,Goldman:2013qla,Tolos2015}, with the~counter-intuitive effect that the more DM the NSs accrete, the~smaller the resulting NSs become, also reducing the maximum mass for a stable NS given an amount of DM~\citep{Sandin2009,Tolos2015}.

Whereas a typical NS has a mass of $\simeq$1.4 $\rm{M}_{\odot}$,~\citep{deLavallaz2010} in recent years, some pulsars were measured at 
$2$ $\rm{M}_{\odot}$ (like PSR J1614-2230 with $1.97 \pm 0.04$ $\rm{M}_\odot$,~\citep{Demorest:2010bx}).

Changing the EoS of the NS or adding DM to it can accommodate such large masses. For~a stiff EoS, observations vs. theory comparison can constrain the EoS and possibly rule)  out NS's exotic matter states (e.g., quarks, mesons, hyperons,~\citep{Schaffner-Bielich2005}), although~some groups report that quark matter does not necessarily lead to a softening (e.g., \cite{Alford:2006vz}).

Concerning the second possibility, in~recent years several authors realized that effects similar to those of exotic states in NS could be obtained in minimally coupled NS matter, admixing DM~\citep{Sandin2009}. 
\mbox{As seen} in Ref.~\citep{Ciarcelluti2010}'s Figure~1 for the $M_{\rm{Radius}}{\rm{vs}}M_{DM}$, where the DM ratio can reach up to 70\%, \mbox{while Ref.~\citep{Goldman:2013qla}} only requires 50\% of DM to reach the 2M$_{\odot}$ mark. More refinedly, the~resulting NS mass is also conditioned by the particle mass of DM~\citep{Li:2012ii,Mukhopadhyay:2015xhs}, so the total NS mass derives from the interplay between DM particle mass and relative DM mass acquired in  the accretion process~\citep{Leung:2011zz}.

In previous studies of mirror matter admixed NSs~\citep{Sandin2009}, degenerate DM~\citep{Leung:2011zz}, or~ADM~\citep{Li:2012ii}, authors all found smaller radii and maximum total masses NSs for increased DM to normal matter ratios. For~example, the~DM admixed quark matter model of Ref.~\citep{Mukhopadhyay:2015xhs} obtained a star mass of 1.95 $\rm{M}_{\odot}$, \mbox{while~\citep{Li:2012ii,Li2013}} obtained NSs with masses $\simeq$2 $\rm{M}_{\odot}$ for DM particles mass of $\simeq$0.1 GeV, in~the case of weakly interacting DM, and~for $m_{\rm{dm}}\simeq 1$ GeV, in~the case of strongly interacting DM. In~\citep{Tolos2015}, NSs, and~White Dwarfs (WDs) objects, admixed with 100 GeV ADM, were studied, resulting in the formation of planets-like objects. The~Compact Objects (COs) study of~\citep{Tolos2015} was continued in~\citep{Deliyergiyev2018} to include, for~NSs, DM  
particle masses spanning 1--500 GeV. They found, for~instance, that decreasing DM particle mass leads to COs increasing mass, as~well as an increase in captured DM by the~COs.

In this paper, the~NSs mass dependence on the environment is examined. Since, the~larger the mass acquired by the NS, the smaller its maximal total mass, it is expected that going towards the GC, or~if we are in the center of dark matter clumps, the~NSs mass must be smaller than that in an environment without DM. This property is used to put constraints on the interaction cross-sections for nucleon-dark matter and DM--DM self interaction.
At the same time, the~decrease of NSs mass towards the GC can be itself used to obtain information on DM's nature. Contrary to the practically untestable proposal of Ref.~\citep{deLavallaz2010}, pointing at NS temperature time evolution correlated with DM accretion, \mbox{our proposal} of mass evolution with DM content is easier to test. The~interest for using probes such as NSs stems from a) the more probable DM--baryon interaction following DM capture because of their very large baryon densities; b) because of the strong gravitational force, after~a DM particle interacts and looses energy it is very improbable it can~escape. 

The paper presents the following organisation. 
In Section~\ref{sec:AccumDMinNS}, we discuss DM accumulation in NS. In~Section~\ref{sec:EstimDMinMW}, we discuss what kind of density profiles must be used to describe the density in our galaxy, taking into account baryon physics and the presence of a black hole (hereafter BH) in the GC. In~Section~\ref{sec:ChangeNSmassDMaccum}, we show that the accumulation of DM reduces the NS maximal total mass, and~this depends from the environment. In~Section~\ref{sec:ConstrainCrossSec}, we discuss how the DM accumulation can have the NS collapse to a BH, and~how this could constrain the nucleon--dark matter cross section. We also find some limitations on the self-interaction of dark matter. Section~\ref{sec:Conclusions}, is devoted to~conclusions.

\section{Accumulation of DM in~NSs}
\label{sec:AccumDMinNS}

An NS containing a non-self interacting (ADM) structure is obtained, as~shown in several papers (e.g.,~\citep{Tolos2015,Deliyergiyev2018}), by~solving the Tolman--Oppenheimer--Volkoff (TOV) equations for an admixture of ordinary matter (OM) with ADM, only coupled by gravity. We refer the readers to~\citep{Deliyergiyev2018} for a description of the equations, and~the way they were~solved. 

In the case of Ref.~\citep{Tolos2015}, the~TOV equations involved ADM admixed with NS and WD material, fixing the DM particle mass equal to 100 GeV, and~for two cases of OM--DM interaction: \mbox{weak interaction}, $y=0.1$, and~strong interaction $y=1000$, where $y$ is defined in~\citep{Tolos2015} as the interaction parameter. The~results of their study showed that the TOV's solutions, in~the case of DM that is weakly self-interacting and non-annihilating, can produce Earth-like masses of Compact Objects (COs) with radii of a few km to a few hundred km, while they obtained, in~the case of DM with strong self-interactions, a~few hundreds km radii, Jupiter-like, COs. They also analyzed the maximum amount DM sustainable by NSs with 2 M$_{\odot}$ and WDs with nominal mass of 1 M$_{\odot}$.
   
\begin{figure}[H]
\centering{}\includegraphics[width=0.98\columnwidth]{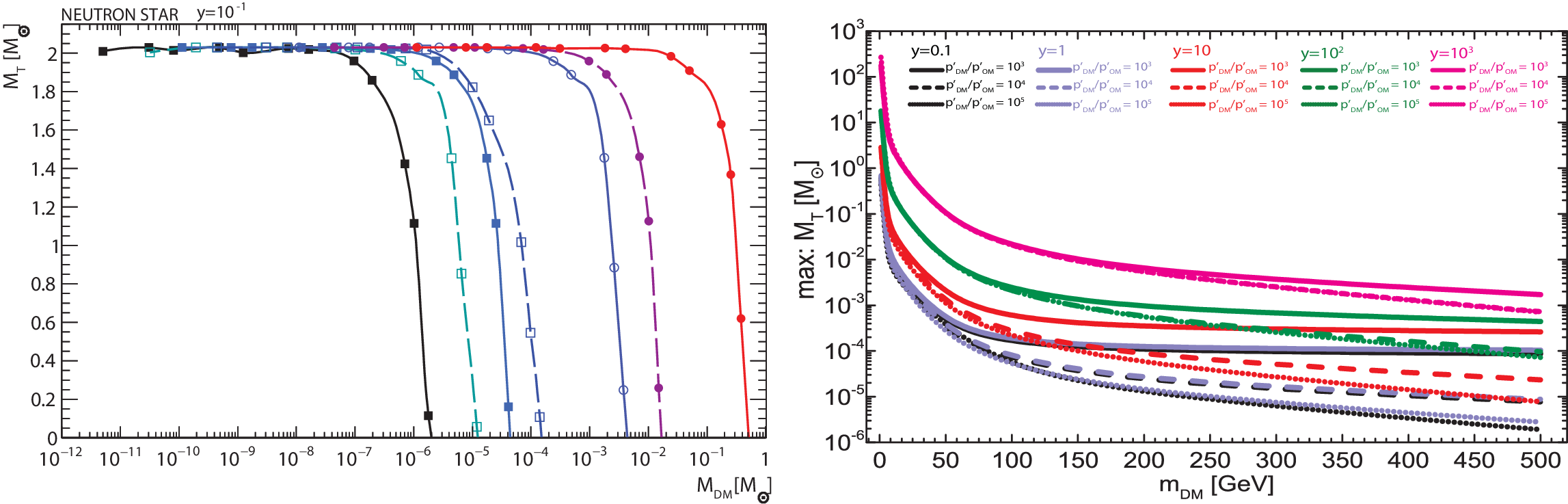} 
\caption{
Left panel: The maximum mass of a stable neutron star (NS) as a function of its dark matter (DM) mass content in the case of $y =10^{-1}$, weakly interacting DM. Each curve corresponds to a given DM particle mass and is built from the last stable point of fixed $p_{DM}/p_{OM}$ curves. They mark how, at~fixed particle mass, the~largest possible NS mass evolves for an increasing DM mass content. The~colour codes the DM particle mass value. From~left to right: $500, 200, 100, 50, 10, 5, 1$ GeV. 
Right panel: the maximum mass of a stable NS as a function of the particle mass for DM, with~various curves for different self interaction cross sections, coded by colour, and~for the DM to ordinary matter ratio, coded by line type.}
\label{fig:NSWD_branches_y01}
\end{figure}

In~\citep{Deliyergiyev2018}, we extended the previous work by considering mass particles equal to $1, 5, 10, 50, 100, 200, 500$ GeV, and~ratio between the ordinary matter (OM) and DM, $p_{\rm DM}/p_{\rm OM}$ in the range $10^{-5}\text{--}10^5$ with step of 10. 
The total COs mass was found to increase with decreasing mass of the DM particle, hence particle masses of strongly self-interacting DM within the range 1--10 GeV were excluded from observed COs. Since we found that more DM is captured in the COs if the DM particle mass is smaller, we constrained from 2 M$_{\odot}$ observations the amount of DM capture. This is shown in Figure~\ref{fig:NSWD_branches_y01} left panel, which plots the total NS mass versus that of DM. Each curve, from~right to left, corresponds to a particle mass of $1, 5, 10, 50, 100, 200, 500$ GeV, respectively. The~right panel of Figure~\ref{fig:NSWD_branches_y01} shows the change of NS maximal total mass as a function of DM particle mass, $y$, and~pressure ratio between DM and ordinary~matter.

Figure~\ref{fig:NSWD_branches_y01} allows us to deduce the precise amount of DM that a formed NS should accrete from their environment to have a decrease in mass, as~we discuss in the~following.

For a Jupiter-like object with mass $\simeq$10$^{-3}$ ${\rm M}_{\odot}$, the~DM content lies in the range $10^{-1}\text{--}10^{-5}$ ${\rm M}_{\odot}$. Those values are predictions of the TOVs equation solution. Given these results, the~natural question is whether natural processes can allow such an amount of DM accreted by a CO. To~answer it, the~amount of DM accreted during the different phases of the CO formation need to be considered. For~the formation of COs of Earth-like, or~Jupiter-like, masses, two phases of DM accretion should be recognized: 
a) the CO collapse phase; 
b) capture by interaction with the CO's nuclei after its collapse phase. In~the case of NSs or WDs, one must distinguish three DM accretion phases: 
a) the star life-time, before~it explodes; 
b) the star to CO collapse phase, 
c) the CO's capture~phase. 

The only simulation, as~far as we know, that models NS accretion of DM from surrounding environment is found in Ref.~\citep{Yang2011}. 
Simplifications such as neglecting the pre-NS phase DM capture~\mbox{\citep{Kouvaris2008,Kouvaris2013}}, or~assuming comparable DM capture between progenitor phase and NS phase~\citep{Kouvaris2010}, allowed some analytical studies to provide NSs accreted DM estimates. Several authors \citep{Kouvaris2008,Kouvaris2010,deLavallaz2010,Yang2011,Kouvaris2011,Guver2014,Zheng2016} have further decomposed the NS phase in three modes of 
DM capture: DM-nucleon scattering; DM-neutron scattering DM orbit decrease;
DM--DM interaction inside the~NS.

The accumulation of DM has been studied by several authors~\citep{Kouvaris2008,Kouvaris2010,deLavallaz2010,Yang2011,Kouvaris2011,Guver2014,Zheng2016},
and happens in several phases~\citep{Guver2014}. In~the first one, the~ambient DM is captured by the NS, when DM scatters with nucleons. In~the second phase, scattering of DM with neutrons produce a decrease in the DM particle radius. In~the third phase, DM interacts with the already captured DM. DM thermalization with neutrons gives rise to the possibility to form a Bose--Einstein condensate, and~DM becoming self-gravitating collapsing and forming a BH~\citep{Guver2014,Kouvaris2013}. The~evolution of the DM number, $N_{\rm dm}$, \mbox{is given by~\citep{Guver2014},}
\begin{equation} 
\label{capt}
\frac{dN_{\rm dm}}{dt}=C_{\rm c}+C_{\rm s} N_{\rm dm}
\end{equation}
 where $C_{\rm c}$ is the capture rate due DM-nucleon interaction, and~$C_{\rm s}$ is the capture rate due to DM self-interactions,given in~\citep{Guver2014}, Equation~(3.8). 
  Following~\citep{Kouvaris2010,Kouvaris2011,deLavallaz2010,Zheng2016} we will neglect the 
the accretion due to~self-interaction.

To calculate $C_{\rm c}$~\citep{Kouvaris2008} assumed a Maxwellian DM distribution. Then the DM orbits intersecting the NS were calculated, and~then the subsample of these loosing enough energy to be captured. In~the case of COs the time-like geodesic equation describing the particle motion in a Schwarzschild metric was used.
The accretion rate can be written as~\citep{Kouvaris2008,Kouvaris2011,Guver2014},
\begin{eqnarray} 
\label{cc}
C_{\rm c}&=&\frac{8 \pi^2}{3} \frac{\rho_{\rm dm}}{m} \left( \frac{3}{2 \pi \overline{v}^2} \right)^{3/2}
 G M R \overline{v}^2 \left( 1-e^{-3 E_0/\overline{v}^2} \right) f \nonumber \\
 & &
=1.1 \times 10^{27} s^{-1}  \left( \frac{\rho_{dm}}{ 0.3 {{\rm{GeV/cm^{3}}}} } \right) \left(\frac{220 
{{\rm{km/s}}}}{v}\right) \left(\frac{{{\rm{TeV}}}}{m} \right)\nonumber \\
&& \left(\frac{M}{M_{\odot}}\right) \left(\frac{R}{R_{\odot}}\right) 
\left( 1-e^{-3 E_0/\overline{v}^2} \right) f
\end{eqnarray}
 where $\rho_{\rm dm}$ is the local DM density, $\overline{v}$ is the average DM velocity, M, and~R the mass and radius of the star, $E_0$ is the DM maximum energy per DM mass 
which can give rise to a capture, and~$E_0 \gg 1/3 \overline{v}^2$, implying $e^{-3E_0/\overline{v}^2} \simeq 0$. $f$ is the fraction of particles undergoing scatterings in the star, and~$f=1$ for $\sigma_{\rm dm} >10^{-45}$ $\rm cm^2$, or~$f=0.45 \sigma_{\rm dm}/\sigma_{\rm crit}$, and~$\sigma_{\rm crit} \simeq 6 \times 10^{-46}$ $\rm cm^2$.

For a typical NS with 1.4 $\rm{M}_\odot$ and a 10 km radius the total mass accreted is given by
\begin{equation} 
\label{Kouv2013}
M_{\rm acc}= 1.3 \times 10^{43} \left( \frac{\rho_{\rm dm}}{\rm 0.3 GeV/cm^3} \right) \left(\frac{\rm t}{\rm Gyr} \right) f \rm GeV
\end{equation}
The previous equation is an underestimate of the accreted mass, since it is not taking into account the accretion by the NS progenitor, which is expected to follow the same order of magnitude compared with that acquired in the NS phase~\citep{Kouvaris2010}, the~accretion due to DM self-interaction~\citep{Guver2014}, and~the fact we use a 2 $\rm{M}_{\odot}$ NS.

There is no typical NS age  $t$. Observation encounters very young pulsars (e.g., the~Crab pulsar), very old ones, e.g.,~>10$^{10}$ years
 (PSR J1518 + 4904, PSR J1829 + 2456), 
or $\simeq2 \rm~Gyr$ (PSR J1811-1736), 
\mbox{as well} as pulsars of intermediate age, of~some $10^8$ years (PSR B1534 + 12
, PSR J0737-3039, PSR J1756-2251, etc.). 
For our calculation, we decided to set 
$t=10~\rm Gyr$.

DM accretion $\simeq 10^{-11}$ ${\rm M}_{\odot}$ was obtained in~\citep{Deliyergiyev2018} for a typical solar neighborhood NS, in~agreement with the  results from ~\citep{Kouvaris2011} and capture rates from Refs.~\citep{Kouvaris2013,Zhong2012,Zheng2016,Guver2014}, but~below the TOV maximum accumulated DM estimates. The~TOV computed maximum accumulated DM  mass can better be reached from accreted DM mass onto NSs wrapped in superdense DM clumps, found in Ultra Compact mini-haloes, close to the GC~\citep{Berezinsky2013}.

However, such high density DM environments are more likely than thought before, \mbox{since, as~pointed} out in many studies~\citep{Berezinsky2003,Ricotti2009,Scott2009,Berezinsky2013,Bringmann2012,Berezinsky2014}, the~halo DM distribution is not homogeneous. Substructures, such as superdense DM clumps (SDMCs), which are radiation dominated era virialized bound DM objects, or~ultra compact mini-haloes (UCMHs), which formed from SDCMs secondary accretion~\citep{Berezinsky2013}, populate the halo. Simulations such as~\citep{Ricotti2009}, or~analytical models~\citep{Berezinsky2013,Berezinsky2014} have been studying SDMC and UCMH~characteristics. 

The spherical, and~ellipsoidal collapse model have been used by~\citep{Berezinsky2013} to determine the characteristics of the SDMC, and~obtained the mass-density relation of SDCMs, and~the overdensity of the perturbation at the horizon crossing time $\delta_H$ (see their Figure~2 
 and Table~1).

\vspace{12pt}  
The annihilation criterion can provide estimates of the clumps center maximum density, and~gives
\begin{equation} 
\rho(r_{\rm min}) \simeq \frac{m_{\rm{dm}}}{\langle\sigma v\rangle (t_0-t_f)},
\label{dmax}
\end{equation}
where $t_0$ is the actual time (13.7 Gyr), $t_f$ the formation time (59 Myr (0.49 Gyr) for non-contracted (contracted) UCMHs~\citep{Scott2009}), $\langle\sigma v\rangle \simeq 3 \times 10^{-26}$ cm$^{3}$/s the thermal cross section, $m_{\rm{dm}}$ the DM particle mass. For~a 100 GeV particle, Equation~(\ref{dmax}) gives $\rho(r_{\rm min})=7.7 \times 10^9$ GeV/cm$^3$, which corresponds to a density $\simeq 2.6 \times 10^{10}$ times larger than the local DM density, and~suggests a DM mass acquired by that NS equal to $\simeq 7.5 \times 10^{-4}$ ${\rm M}_{\odot}$.

\section{Estimating the Quantity of DM in the~MW}
\label{sec:EstimDMinMW}

That DM density increases toward the galaxy center is known, although~its exact density profile is not, and~in particular whether the profile is cored, as~seen in many dwarf spiral galaxies, or~cuspy. Probing and determining the MW structure with current data has not yet allowed to disentangle between such profiles~\citep{Ullio2010,Weber2010}.
Several spherical averaged profiles have been proposed.
N-body simulations predict cuspy profiles for all galaxies (e.g.,~\citep{Navarro1997}). The~Navarro--Frenk--White profile~\citep{Navarro1997} has an inner logarithmic slope $\beta=-1$, while more recent simulations~\citep{Stadel2009,Navarro2010} shows a flattening going towards the galactic center, to~a minimum logarithmic slope of $\beta=-0.8$ at radius of 120 pc~\citep{Stadel2009}, resembling to the so-called Einasto profile, given by
\begin{equation}
\rho=\rho_{{-2}}{{\rm e}^{-2\,{\frac {1}{\alpha} \left[  \left( {\frac {r}{r_{
{-2}}}} \right) ^{\alpha}-1 \right] }}},
\end{equation}
where $\alpha$ controls the density profile's degree of curvature, such that smaller $\alpha$ values correspond to cuspier profiles, $r_{-2}$ sets the radius for which $\frac{d \ln{\rho}}{d \ln{r}}=-2$, and~where $\rho_{{-2}}$ is the corresponding density. The~Einasto profile has three free parameters. 
To obtain a realistic halo density profile in the MW, \mbox{we must} fix the quoted parameters. This can be done in different ways as it was discussed in~\citep{DelPopolo2020}.

However, the~previous discussed profiles, including the Einasto profile, are not taking into account baryon physics.
The last has two effects on the density profile: 
a) steepens it in the so called adiabatic contraction~\citep{Gnedin2004,Gustafsson2006,Pedrosa2009,Duffy2010}; 
b) may flatten the profile due to supernovae feedback or similar feedback effects~\citep{DelPopolo2010a, DiCintio2014,DelPopolo2016a,DelPopolo2017}.
Semi-analytic models~\citep{DelPopolo2010a,DelPopolo2016a}, and~hydrodynamic simulations~\citep{DiCintio2014} show a dependence of the inner slope from mass, with~flatter profiles in dwarf galaxies and cuspy ones, 
$\beta \simeq -1$ for galaxies with mass similar to that of the Milky Way. 
The flattening of the density profile in dwarf galaxies is due to the efficient outflows of gas due to supernovae feedback, or~interaction of dark matter with baryons through dynamical friction~\citep{DelPopolo2009,DelPopolo2016a} while in more massive galaxies the deepening of the potential due to the presence of more stars, make the outflow mechanism inefficient with the result of having cuspier profiles.
Then for galaxies as massive as our MW adiabatic contraction has a predominant effect.
Ref.~
 \citep{Prada2004} made the correction to a density profile due to adiabatic contraction (see their Figure~1), showing that in the inner 3 kpc, where baryons are dominating, it is not sensible to use the DM-only profiles given by simulation. This is confirmed by the smoothed particle hydrodynamics (SPHs) simulations, and~the so called DC14 profile~\citep{DiCintio2014}. In~the case of an halo as massive as our MW, the~density profile's inner logarithmic slope, $\beta \simeq -1.2$ is even steeper than that of the NFW model. 
A similar result was obtained by~\citep{Pedrosa2009}, who obtained $\beta \simeq -1$ for DM-only simulations and $-$1.25 in presence of baryons. 
In Figure~1 of~\citep{DelPopolo2020}, we plotted some Einasto profiles, and~that of~\citep{DiCintio2014}, and~used it to calculate the DM acquired by NSs. That calculation is however limited, as~previously reported, because~Einasto's profile are DM-only profiles.
We need to take account of the role of the baryons, and~also the fact that close to the BH the profiles are expected to be very~cuspy.

Since the profile of \citep[]{DiCintio2014} (hereafter DC14) shows very good agreement with, and~describes very well that of observed galaxies, it can be considered to provide a good description of the real profile close to the galactic centre. This profile yields a density of $\simeq10^{11} \rm$ GeV/cm$^3$ at $10^{-5}$ pc. Indeed, \mbox{many authors} agree that the density close to the galactic centre is of the order of, or~even much larger than, the~DC14 profile
 (i.e.,~\citep{deLavallaz2010,Bertone2005a,Gondolo1999}). 

Furthermore, in~addition to the central cusp, haloes can foster other high density substructures: DM density spikes are expected by Ref.~\citep{Sandick2018} in the GC, while for Lacroix~\citep{Lacroix2018}, in~cuspy outer halos, a~spike with few tens of parsecs or smaller radius is not excluded. A~recent examination of spikes can be found in Ref.~\citep{Bennewitz2019}, also providing references on the discussion around their existence (i.e.,~\citep{Fields2014}). 
Efficiency of accretion is also an important factor. Much larger NS accumulation of DM than found in our past work (i.e.,~\citep{Deliyergiyev2018,DelPopolo2020}) is predicted from Ref.~\citep{Bertone2005a} combined with, e.g.,~PSR B1257 + 12 orbital dynamics~\citep{Iorio2010}, yielding DM accretion over NS mass ratio up to 10\%, reaching environments effects in agreement with DC14~\citep{DiCintio2014}. The~picture of DM accretion is further complicated by astrophysical phenomena, located near the GC on sub-parsec scales, including star gravitational scattering, supermassive BH capture and supermassive BH formation's enhanced central density~\citep{Bertone2005a,Gondolo1999}. Taking those into account leads to improved density profiles~\citep{deLavallaz2010,Bertone2005a}.

\begin{figure}[H]
\centering{\includegraphics[width=1\columnwidth]{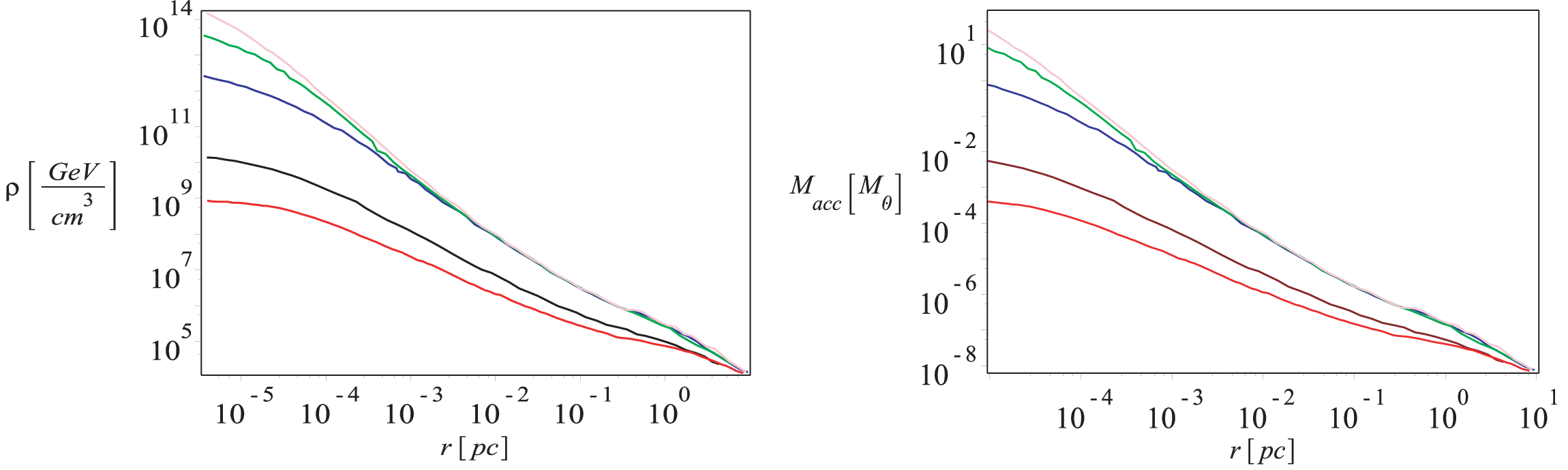}
}
\caption{
{\textbf{Left Panel}}:~\citep{Bertone2005a}'s density profiles. 
\textbf{Right panel}: accreted mass obtained from the density profiles in the left panel using Kouvaris formula. 
Pink line is the reproduction of the top line from Figure~1 of~\citep{Bertone2005a}  (yellow line), while the green and blue line correspond to the central  (green)  and bottom (blue) lines  in~\citep{Bertone2005a}. 
The black, and~red lines belong to the same family but correspond to the less steep density profile.
}
\label{fig:einasto}
\end{figure}

For these reasons, in~this paper we use more physical profiles than~\citep{DelPopolo2020}, like the~\citep{Bertone2005a} profiles. In~Figure~\ref{fig:einasto}, left panel, we plotted some of the density profiles. Figure~\ref{fig:einasto}'s right panel displays the accreted DM corresponding to each density profile, computed from Equation~(\ref{Kouv2013}), the~\citep{Kouvaris2013} formula. The~top line in Figure~1 of~\citep{Bertone2005a}  (yellow line)  is reproduced here as the pink line, while the green and blue line correspond to the central  (green)  and bottom  (blue)  line in~\citep{Bertone2005a}. The~black, and~red line belong to the same family but are less steep. Note, that in Figure~\ref{fig:einasto}, right panel, at~$10^{-5}$ pc, some of the curves give a very large accretion, so we consider the validity of the curves in the region  $>10^{-3}$, for~which all the curves give an accretion smaller than 1\%.

\unskip

\section{Change of NS Mass Due to DM~Accumulation}
\label{sec:ChangeNSmassDMaccum}

\begin{figure}[H]
\centering{
\includegraphics[width=1\columnwidth]{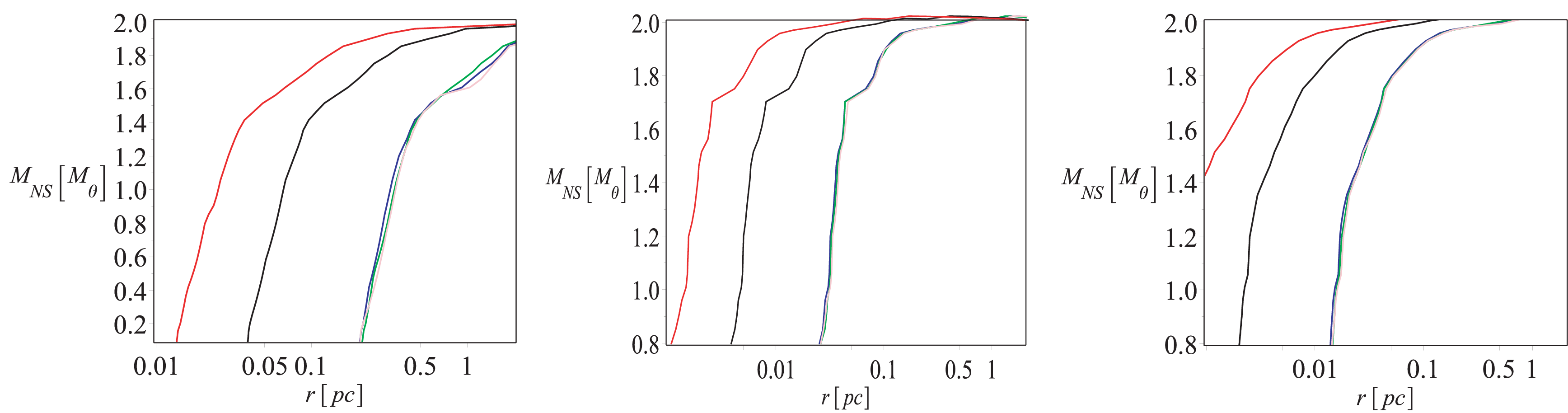}
}
\caption{\label{fig:M_NS_45} The NS mass changes due to DM accumulation in terms of distance from the galactic center, for~cross section $10^{-45}$, particle mass $m_{\rm{dm}}=500$ (\textbf{left panel}), 200 (\textbf{central panel}), 100 (\textbf{right panel}), and~for $y=0.1$. The~colours correspond to those used in Figure~\ref{fig:NSWD_branches_y01}.\label{fig:NSmassChangeDMaccum}
}
\end{figure}

Now, we can calculate how the accumulation of dark matter influence the NS mass. 
In order to do this, we notice that since $M_{\rm acc}$ of Figure~\ref{fig:einasto} is equal to $M_{\rm DM}$ in Figure~\ref{fig:NSWD_branches_y01}, combining those figures together,  we get the NS total mass vs radius relation, $M_{\rm T}-r$. In~fact Figure~\ref{fig:NSWD_branches_y01} and Figure~\ref{fig:einasto} give a relation 
$M_{\rm T}-(M_{\rm DM}=M_{\rm acc})-r$. 
This procedure is repeated for the particle masses 500, 200, and~100~GeV, at~$y=0.1$, and~plotted in Figure~\ref{fig:NSmassChangeDMaccum}. 
The plot shows that the pink, green, and~blue line, the~steepest line in Figure~\ref{fig:einasto} (left panel), give similar results concerning the decrease in mass of NSs: these NS reach  maximal masses of the order of 0.2 M$_{\odot}$ around 0.3 pc, for~particle masses of 500 GeV. The~black, and~red line which correspond to less steep profiles in Figure~\ref{fig:einasto} (left panel) reach 0.2 M$_{\odot}$ at smaller radii. 
\mbox{Here, we} want to recall that the theoretical smaller mass for a NS is of the order of 1 M$_{\odot}$. 
\mbox{This means} that with the accretion of mass, the~star will not be longer stable. 
The central and right panel of Figure~\ref{fig:NSmassChangeDMaccum}, gives similar information for the cases the DM particles have a mass of 200, and~100~GeV.

\unskip

\section{Constraints on the Cross~Section}
\label{sec:ConstrainCrossSec}

As written in the introduction, DM captured by a NS thermalizes and concentrates in a small central region of radius $r_{\rm th}=\rm 220$ cm $(\frac{GeV}{m_{\rm{dm}}})^{1/2} (\frac{T_c}{10^5})^{1/2}$, where $T_c$ is the central temperature. \mbox{Then DM} becomes self-gravitating and forms a Bose--Einstein Condensate (BEC) of radius $r_{\rm BEC} \simeq 1.6 \times 10^{-4} (\frac{\rm GeV}{\rm m_{\rm{dm}}})^{1/2}$ cm for a 10 TeV particle. If~the DM acquired is larger than
\begin{equation}
M>8 \times 10^{27} \rm GeV \left(\frac{GeV}{m_{\rm{dm}}}\right)^{3/2}
\end{equation}
a BH with mass of
\begin{equation} \label{critic}
M_{\rm crit}=\frac{2 M_{\rm pl}^2}{\pi m_{\rm{dm}}} \sqrt{1+\frac{\lambda M_{\rm pl}^2}{32 \pi m_{\rm{dm}}^2}}
\end{equation}
is formed~\citep{Kouvaris2013}, where $\lambda$ is the self-interaction term. 
The mini BH can consume the star, and~destroy it. Hawking radiation counteracts this effect. 
The BH evolution is given by
\begin{equation}
\frac{dM}{dt}=\frac{4 \pi \rho_c G^2 M^2}{c_s^3}-\frac{f}{G^2 M^2}
\end{equation}
where $c_s$ stands for the sound speed inside the core, while f gives a dimensionless radiation pressure factor, depending on some effects. Accretion wins Hawking radiation if $M> 5.7 \times 10^{36}$ GeV. We will see this~later.

In the absence of self-interactions, the~bosonic Chandrasekhar bound scales as $N_{Chand}\approx (M_{pl}/m_{\rm{dm}})^{2}$. For~non-interacting fermions, the~Chandrasekhar bound instead scales as $N_{Chand}\approx (M_{pl}/m_{\rm{dm}})^{3}$ due to the degeneracy pressure. As~a result, old NSs could not have accumulated enough non-interacting fermionic DM to initiate gravitational collapse. If~bosonic DM has even a very small repulsive self-interaction, then the bosonic Chandrasekhar bound could also scale as $(M_{pl}/m_{\rm{dm}})^{3}$, leading to elimination of any bound on DM. However, if~fermion DM has an attractive self-interaction, then this force can compensate for the Fermi degeneracy pressure, allowing for DM collapse to BH Ref.~\citep{Bramante:2013nma, Bramante:2014zca}. For~bosonic DM, a~purely attractive interaction may be expected to lead to a vacuum instability. However,~for fermionic DM, a~Yukawa interaction could lead to a consistent attractive self-interaction. \mbox{It is thus} of interest to determine the extent to which fermionic DM with Yukawa self-interactions can be constrained by observations of old NS, and~in particular, if~the constraints have implications for the parameter space relevant for astronomical evidence for self-interacting DM~\citep{Bramante:2013nma}. Projecting recent results on fermionic systems from solid state physics Ref.~\citep{PhysRevLett.125.196403}, one may speculate that attractive DM effects may emerge from a strong repulsion in the fermionic DM~systems.

NS constraints have mostly been applied to bosonic DM, which has no Fermi degeneracy pressure to obstruct gravitational collapse. In~this paper we employ the fermionic gas with the repulsive vector interactions.  For~small $y\ll 1$, the~EoS will be similar to an ideal Fermi gas while for large $y\gg 1$ the EoS will be mostly determined by the interaction term, unless~$z$ becomes small enough so that the EoS becomes dominated by the ideal gas term.

However, old NSs have been observed, so they were not consumed by BHs. This can put constraints on some types of~ADM.

\begin{figure}[H]
\centering{\includegraphics[width=0.8\columnwidth]{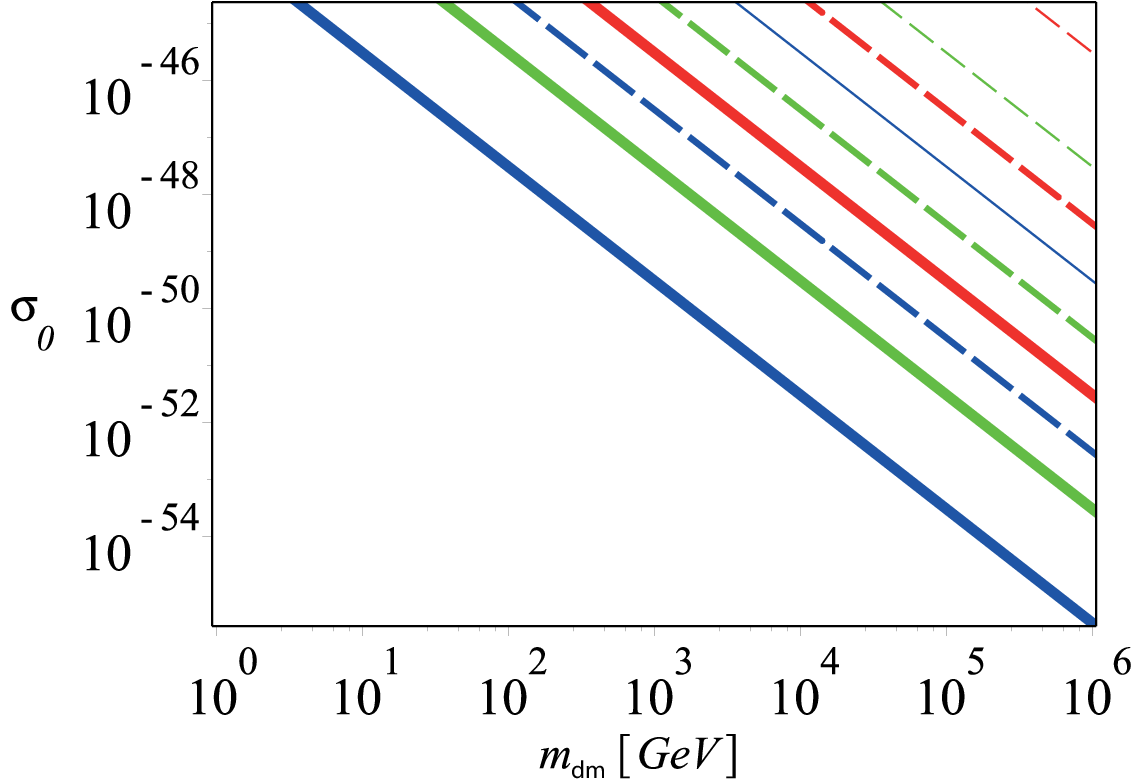}
}
\caption{
Dark matter nucleon cross section obtained from the collapse of NSs to BHs due to accumulation of DM.
The red lines correspond to $\sigma_0$, and~$m_{\rm{dm}}$, giving rise to a collapse in $10^6$ years. 
The thicker the line the higher the density from bottom (solid lines) to top (short dashed lines)---in decreasing thickness and therefore density: $10^{11}$, $10^{8}$, $10^{5}$. 
 The green lines case giving rise to a collapse in $10^8$ years. The~blue lines correspond to case of $10^{10}$ years.} 
\label{fig:DMnCrossSection}
\end{figure}

Should accretion take over Hawking radiation, the~thus created BH would entirely swallow the NS. This is an interesting possibility for several reasons. The~first 
resides in its providing an additional solution to the unexplained observed gamma ray bursts, distinct from the proposed NS coalescence with another CO. Secondly, one can put constraints on some quantities. 
For a given time of capture of DM, and~a DM particle mass, it is possible to determine the  required cross-section that can lead to the accretion of a mass equivalent to $M_{\rm crit}$, amount that will induce the collapse of the NS to a BH. In~this way, one can pick out on
the $\sigma_0-m$ ($\sigma_0$ is the nucleon dark matter cross section, indicated before with $\sigma_{\rm dm}$) plane where this could occur. For~a value of the interaction parameter $\lambda$ such that $M_{\rm crit} \simeq M_{\rm Chandrasekhar} \simeq M_{\rm Pl}^3/m^2$, as~in~\citep{deLavallaz2010}, we consider three times for accretion $10^6$, $10^8$, \mbox{and~$10^{10}$ years}, and~the following densities $10^{11}$, $10^{8}$, $10^{5}$ $\rm GeV /cm^3$.
We should recall that there is a limiting effective cross section related to the cross section of the NS ($\sigma_0^{\rm max}= 2 \times 10^{-45} \rm{cm}^2$), then we have a  minimum value for the DM particle mass below which the fixed accretion time is too short for the star to accrete $M_{\rm crit}$.
In Figure~\ref{fig:DMnCrossSection}, the~red lines correspond to $\sigma_0$, and~$m$, giving rise to a collapse in $10^6$ years. From~bottom (solid red line) to top (short dashed lines), the~lines represent densities equal to  $10^{11}$, $10^{8}$, $10^{5}$. 
 The region of interest of the direct detection experiments are obtained only for masses in the order of 100 GeV, long accretion times ($10^{10}$ years),and high densities ($10^{11}$ GeV). For~higher dark matter masses the cross sections become smaller, and~this because the mass of the particle is larger, and~we need a smaller number of particles to reach $M_{\rm crit}$.
The green lines case is similar to the previous one, but~now the collapse time is $10^8$ years. The~blue lines are the case $10^{10}$ years.

\begin{figure}[H]
\centering{
\includegraphics[width=0.49\columnwidth]{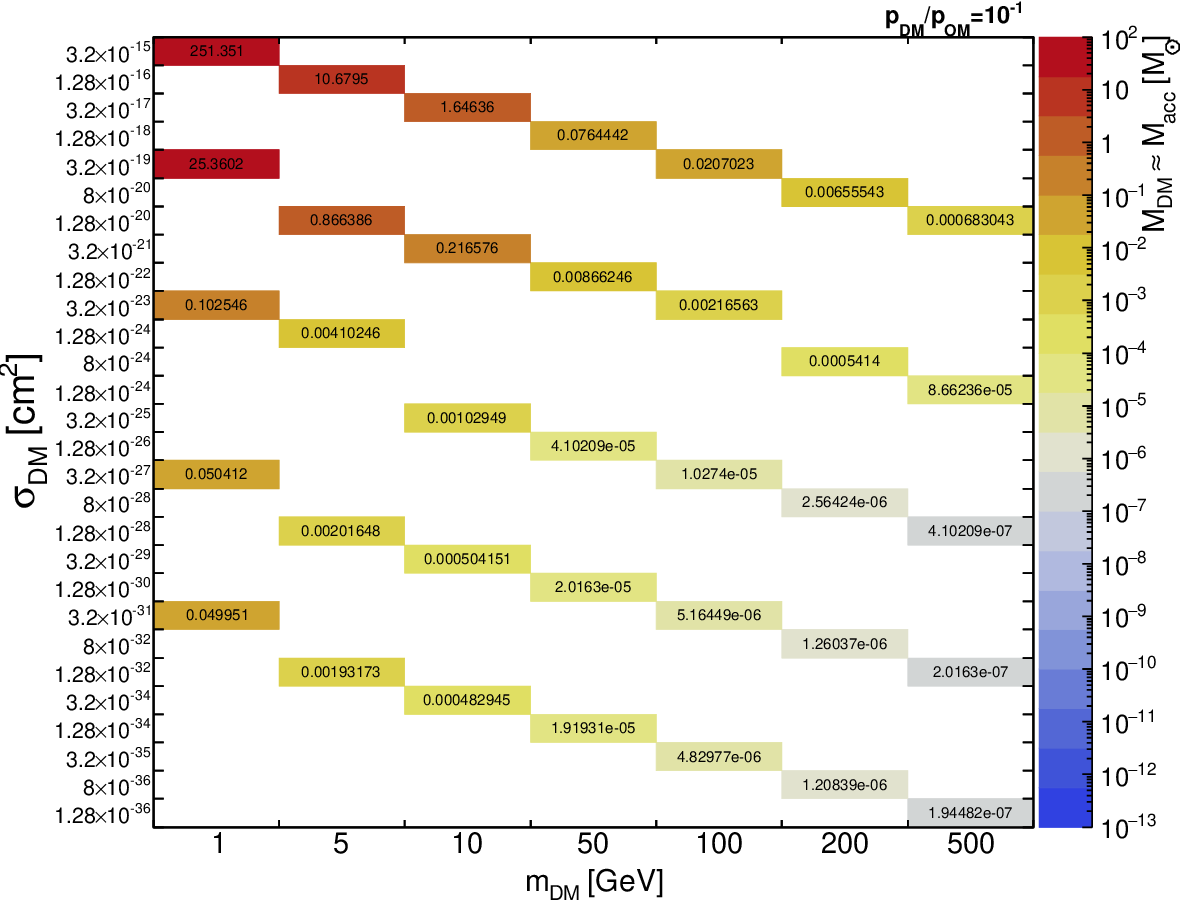}
\includegraphics[width=0.49\columnwidth]{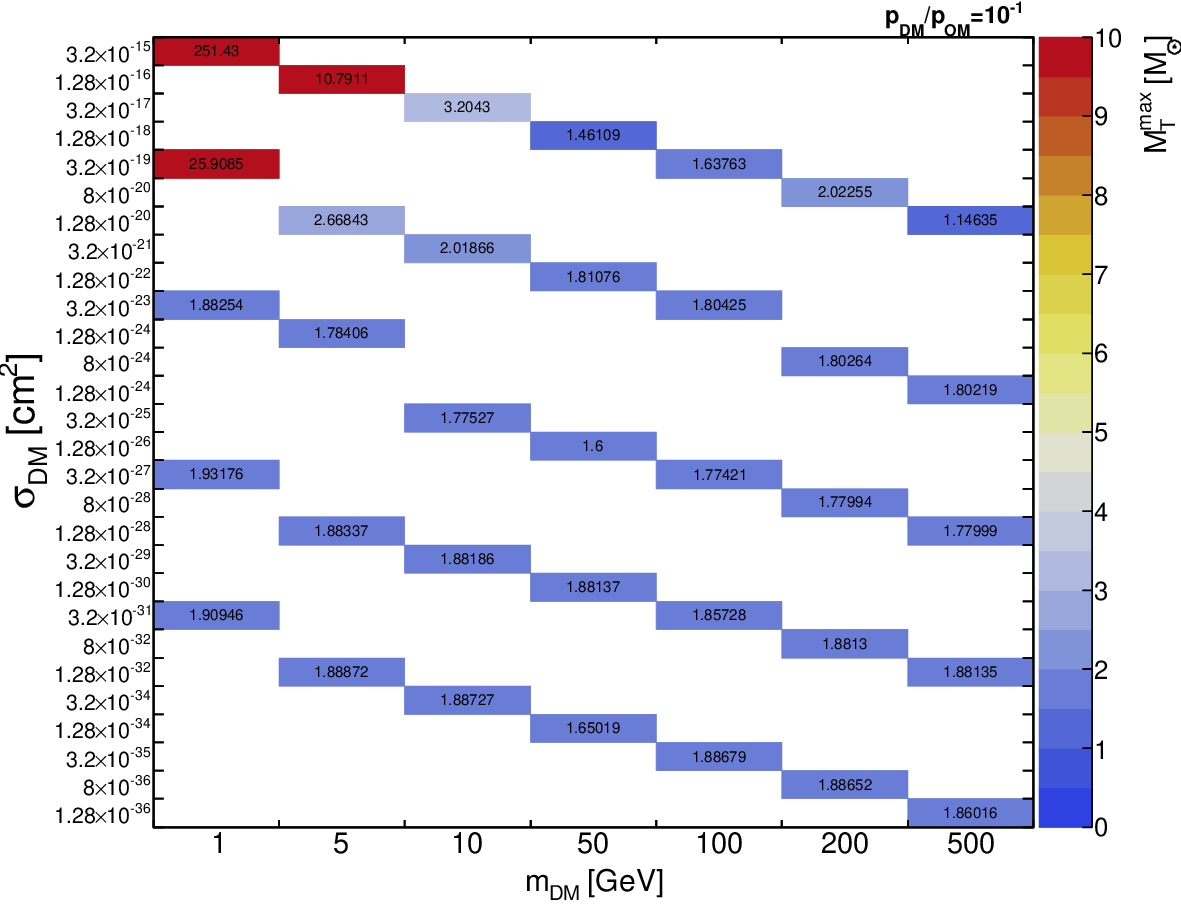}
}
\caption{
Self-interaction dark matter cross section, in~the case $p_{\rm DM}/p_{\rm OM}=0.1$. 
The red bins indicate forbidden regions of the parameter space, where the accumulated DM mass is much greater than ordinary mass of the NS. Therefore, the~maximal total mass of NS goes over the conventional mass limit of 2 solar masses.
The slopes indicating the different DM interaction parameter, $y=1000, 100, 10, 1, 0.1$ from top to bottom.
}
\label{fig:SelfIntDM}
\end{figure}

The left panel of Figure~\ref{fig:SelfIntDM}, which only plots the $p_{\rm DM}/p_{\rm OM}=0.1$ case, provides another constraint.
The red bins indicate forbidden regions of the parameter space. This become more contrast on the right panel where we show the maximal total mass of NSs. It follows that for $m_{\rm{dm}}=1$ GeV, NSs with $y>10$ are forbidden, they are too massive. Similarly for $m_{\rm{dm}}=5$ GeV and $y>100$. 
In the case, $m_{\rm{dm}} \geq 10$~GeV, the~mass of NSs fall in the acceptable~region. 

Before concluding, we want to answer a legitimate question: can observations determine the NSs mass change, and~how many NSs can we see in the GC? This question is all the more pressing that from the galaxy's inner $30'$, only six NSs have been detected~\cite{Eatough2013,Mori2013,Kennea2013,Shannon2013}, including the transient magnetar J1745-2900~\cite{Eatough2013,Mori2013,Kennea2013,Shannon2013}, located 0.1 pc from the GC, while up to thousands of pulsars are theoretically expected in the GC~\cite{Pfahl2003,Wharton2011,Chennamangalam2013}.

Pulsar emission suppression mechanisms, more complex scattering models, stellar population synthesis arguments~\cite{Bower2018}, or~even hyper-strong interstellar scattering~\cite{Lazio1998}, have been proposed to understand this ``missing pulsars''~problem. 

Increased observation power can also hope to bridge the gap between expected and accounted for Pulsars. The~projected capacities of Square Kilometer Array (SKA) allow us to project that the instrument should detect, for~instance, a~5-ms Millisecond Pulsar (MSP) with luminosity $L_{1000} \simeq 0.7\, \rm mJy \, kpc^2$, spectral index $\alpha=-1$ and signal/noise ratio of $S/N = 10$, at~the GC~\cite{FaucherGiguere2010}. Some authors estimate that only a few pulsar--BH binaries should be detected in the inner parsec near the GC~\cite{FaucherGiguere2010}, others assess their conservative upper limit at $\simeq$200 \cite{Chennamangalam2013}, while up to 52 canonical pulsars~\citep{Rajwade2017}, and~up to 10,000 MSPs are predicted for detection by the next generation Very Large Array (ngVLA) and SKA surveys~\citep{Murphy:2018vxa,Keane2015}. The~projected improvement of sensitivity at high frequencies by a factor of 10 of the ngVLA~\citep{Murphy:2018vxa} is expected  to not only dramatically improve GC neighbourhoor pulsar detection, but~is also expected to offer an unparalleled probe of BH physics and General Relativity, in~the line with the first image of the M87 BH by the Event Horizon Telescope (EHT) \citep{Luminet2019}.

Pulsar detection is not the final word to settle the mass change question: their mass also need to be measured. Several techniques of mass measurements are already known~\cite{Engineer1998, Watts:2016uzu}, which can be completed with a recently proposed one, constraining nuclear and superfluid EoS models from pulsar glitch data~\cite{Ho:2015vza}. 
Masses and radii precise measurements, needed to pin down NS composition, \mbox{are expected} from the future instruments Athena~\citep{Barcons:2012zb,Athena:2014cdf}, eXTP~\citep{Watts:2018iom}, SKA~\citep{Konar:2016lgc} and NICER~\citep{NICER:2012} and should help settle the mass change~question.

\section{Conclusions}
\label{sec:Conclusions}

In this paper, we used the~\citep{Bertone2005a} profiles, and~the results of~\citep{Deliyergiyev2018} extended to a larger set of self-interaction cross section, $y$, to~determine the mass accreted by NS at different distance from GC. While in the Sun neighborhood the quantity of DM acquired is small, going towards the GC it becomes large, and~can bring to change of the structure of the NS. One of the many consequences of DM accumulation, already described in~\citep{Deliyergiyev2018}, is a decrease in mass of the NS when acquiring dark matter. This was already implicit in 
\citep{Deliyergiyev2018}, but~we showed how the mass decreases going towards the GC. This effect by itself can be used as a probe on DM. If~DM is of the ADM type, one should observe the quoted decrease of NS mass going towards the GC. Moreover, the~acquisition of DM, when it reaches values close to the Chandrasekhar mass, renders unstable the NS and collapses it into a BH. Such an event can provide, from~one side,  an~additional solution to the universe's unexplained observed gamma ray bursts distinct from the proposed NS coalescence with another CO. It can moreover constrain the cross section, as~shown.


\authorcontributions{All authors contributed equally. All authors have read and agreed to the published
version of the manuscript.} 

\funding{MLeD acknowledges the financial support by Lanzhou University starting fund and the Fundamental Research Funds for the Central Universities (Grant No.lzujbky-2019-25).} 

\acknowledgments{
 M.D. thanks Xin Wu and UNIGE administration for their support in this research during the COVID-19 quarantine measures.}
\conflictsofinterest{The authors declare no conflict of interest.}


\reftitle{References}

%

\bibliographystyle{apsrev4-1}
\bibliography{biblioNS,old_MasterBib2}


\end{document}